\documentclass[a4paper,11pt]{article}

\usepackage{jinstpub} 
\usepackage{graphicx}
\usepackage{xcolor}
\usepackage{caption}
\usepackage{subcaption}
\usepackage[numbers,sort&compress]{natbib}

\title{Imaging of muon track in CsI(Tl) crystal with single photon sensitive camera}

\author[a,b,c,1]{Zhimin Wang,\note{Corresponding author.}}
\author[a,b]{Min Li,}
\author[a,b]{Diru Wu,}
\author[a,c]{Jinchang Liu,}
\author[a,c]{Yongpeng Zhang,}
\author[a]{Xiangcheng Meng,}
\author[a,b]{Caimei Liu,}
\author[a,b]{Changgen Yang}

\affiliation[a]{Institute of High Energy Physics, Chinese Academy of Sciences, Beijing 100049, China}
\affiliation[b]{University of Chinese Academy of Sciences, Beijing 100049, China}
\affiliation[c]{State Key Laboratory of Particle Detection and Electronics, Beijing 100049, China}
\emailAdd{wangzhm@ihep.ac.cn}

\abstract{As a novel approach on visual photon imaging by a single photon sensitive camera and PMTs, this work is trying to measure and identify muon tracks from the 2-D images of CsI(Tl) crystal (scintillator detectors). It is possible that muon tracks can be seen directly with a good signal-to-noise ratio neither with further amplification nor external light, which provides an evolution method for particle measurement in the photon-starved regime of scintillation detectors. The setup of the crystal and camera testing system and the identification algorithm of muon track will be discussed in detail including the system calibration, identification model, signal-to-noise ratio, muon track confirmation, and an expectation on further improvements and applications.
}

\keywords{photon detectors, scintillator detector, imaging, single photon, camera, muon track}

\arxivnumber{1234.56789} % only if you have one

\date{Received: date / Accepted: date}

\begin{document}
\maketitle
\flushbottom

\section{Introduction}
\label{sec:1:intro}

Vertex and track reconstruction are critical for most particle physics experiments, such as studies on neutrino, dark matter, and others. There is a long list of related technologies including but not limited to emulsion film\cite{emulsion-opera-2011,emulsion-MORISHIMA201519,emulsion-2016}, cloud chamber\cite{cloud-chamber-LONGAIR201455}, bubble chamber\cite{pico-bubble-chamber-Bressler_2019}, spark chamber\cite{spark-RUTHERGLEN19641}, multi-wire proportional chamber\cite{MWPC-BALL1979263}, TPC\cite{TPC-Hilke_2010}, Si strip\cite{Si-strip-CACCIA1987124} and Si pixel\cite{si-pixel-2004} etc. In the case of photon-based detection, in particular, PMT or SiPM is the commonly used sensor for timing, intensity, and crude spatial reconstruction, such as JUNO\cite{junodetector,juno-PMT-2022}, Darkside\cite{dark-side-Aalseth_2020}, JUNO-TAO\cite{juno-tao-2020}, SNO$+$\cite{SNO-DUNGER2019162420,SNO-Albanese_2021}, and DUNE\cite{DUNE:2020lwj} etc., where computer algorithms are further used to have a better reconstruction on the vertex or track\cite{JUNO-Rec-Marcel,JUNO-Rec-QLiu,JUNO-Rec-Zhen}.

Recently, many efforts are focusing on photon imaging-related projects following the new development of sensors, where the critical challenges are the need for high spatial resolution over large volumes\cite{BATTAT20161} and better effective signal-to-noise ratio under the photon-starved regime.

For many years classical emulsion film radiography is being replaced by digital detector imaging, especially in medical applications due to faster and more reliable diagnostics and computed tomography and tomosynthesis capabilities\cite{Schumacher2016PhotonCA}.
The single photon counting X-ray CCD camera spectrometer is used in laser-plasma interaction experiments as a simple tool to study the K-shell X-ray generation. A CCD detector enables the spectrum of the impinging X-ray radiation to be obtained without further dispersive devices\cite{Single-photon-CCD-2014}.
Among the imaging systems used for thermal neutron imaging worldwide, the most prevalent configuration is CCD camera based\cite{MOR2021165632}.
Single-photon light detection and ranging (lidar) offers single-photon sensitivity and picosecond timing resolution, which is desirable for high-precision three-dimensional (3D) imaging over long distances\cite{3Dimaging-45km-Li:20}.
Single image 3D photography enables viewers to view a still image from novel viewpoints\cite{jampani2021slide}.
Some good sensors are developed too, such as SPC3\cite{SPC-v2}, a single photon counting camera based on a 2-D imaging array. 
A small, high resolution, high signal-to-noise GEM-based TPC with a 2-D CCD readout designed to provide a benchmark for background discrimination and directional sensitivity that could be used for future optimization studies for directional dark matter experiments \cite{PHAN201682,Deisting2021CommissioningOA}. A skipper CCD was also developed for very low noise and directly measured a muon track through ionization inside the sensor\cite{Skipper-CCD-osti_1839558}. 

But, generally, it is not suitable to directly image of vertex or track in case of a starved-photon regime and uniform angular distribution of the photons\cite{BATTAT20161,LS-imaging-PhysRevD.97.052006}.
Photography by CCD or other technologies, in particular single photon imaging, provides another new possibility, such as our previous study for particle imaging by event\cite{camera-Wang:2022xgb}.

In this article, we will try to have a further detailed check on the imaging of muon track in CsI(Tl) crystal with a single photon sensitive camera and PMTs. Sec.\ref{sec:1:system} will introduce the system setup and calibration. Sec.\ref{sec:1:muon} will discuss the expected features of muon tracks, measurements, and track surveys. Sec.\ref{sec:1:dis} will provide further discussions on the results, possible improvements of the system, and further expectations. And a short summary is in Sec.\ref{sec:1:summary}.

\section{CsI(Tl) crystal with camera}
\label{sec:1:system}

\subsection{Setup}
\label{sec:1:system:2:setup}

\begin{figure*}[!ht]
    \centering
    \includegraphics[scale=0.5]{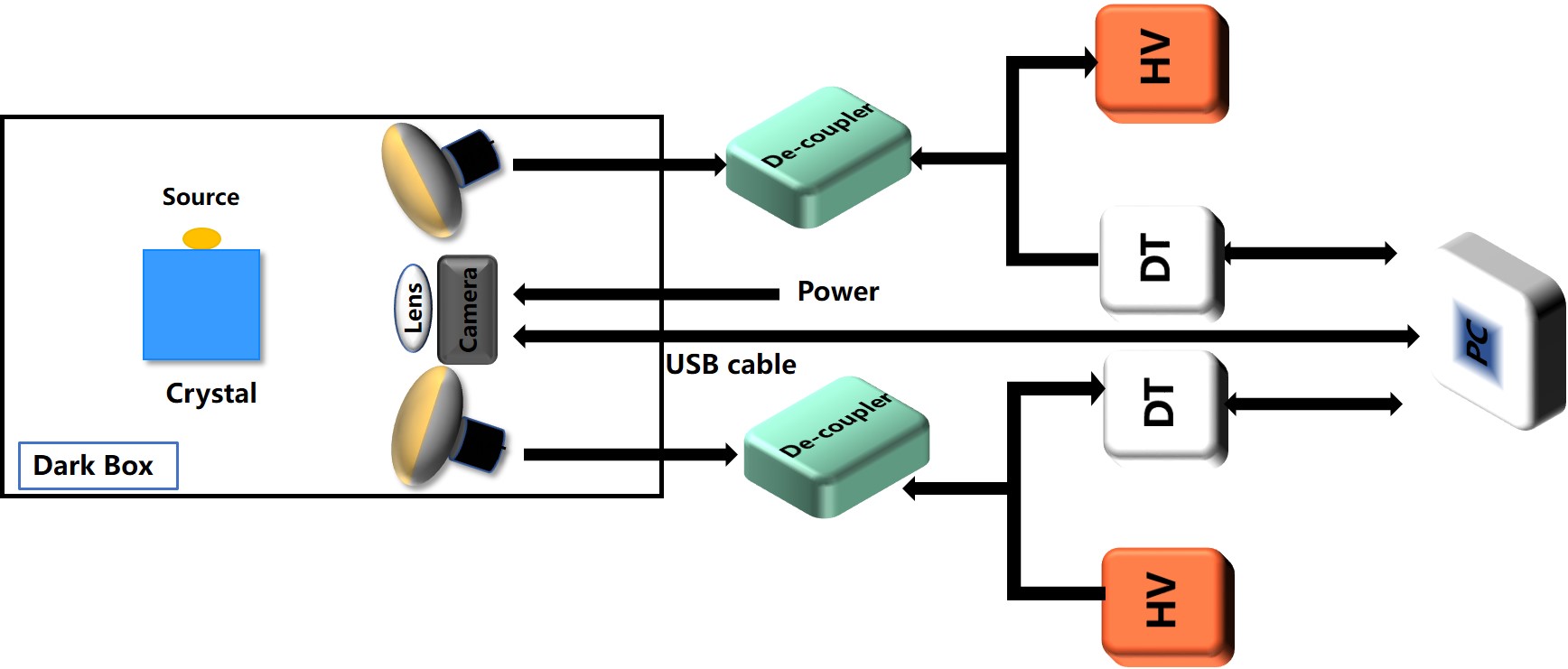}
    \caption{Layout of the imaging measurement system.}
    \label{fig:system}
\end{figure*}

An imaging system, as in \cite{camera-Wang:2022xgb}, is set up with a single photon sensitive camera of ORCA-Quest qCMOS C15550-20UP, which is a new product of Hamamatsu Photonics\,\cite{HPK-ORCA-QUEST}. The detailed layout of the system is shown in Fig.\,\ref{fig:system}. The output of the camera will save in tif format with 16\,bits of each of the 4096(H)$\times$2304(V) pixels and the volume of each photo is around 16\,MB. The CsI(Tl) crystal ($7.5\times7.5\times15\,cm^3$) is located in front of the camera and the two 3-inch PMTs, where the distance between them can be adjusted. An alpha source of $^{241}Am$ is used and put on the top surface of the crystal. The two 3-inch PMTs are used to calibrate and monitor the signal intensity of the crystal, the coincidence of which is used as a trigger of the CAEN DT5751 (1\,GS/s with 1\,V p-p dynamic range, \cite{CAEN-dt5751}) for waveform data taking. The threshold of each 3-inch PMT is set to around 1\,p.e.\,(photon-electron). 
The maximum rate of the data-taking system is limited by the DT5751, which is generally lower than 100\,Hz with data saving of four channels and 10000\,samples of each channel. Here the window length of the waveform recording is set to 6\,\textmu s (6000\,samples/waveform), and the maximum data-taking rate is around 70-80\,Hz.

\begin{figure*}[!ht]
    \centering
    \begin{subfigure}[c]{0.45\textwidth}
	\includegraphics[width=\linewidth]{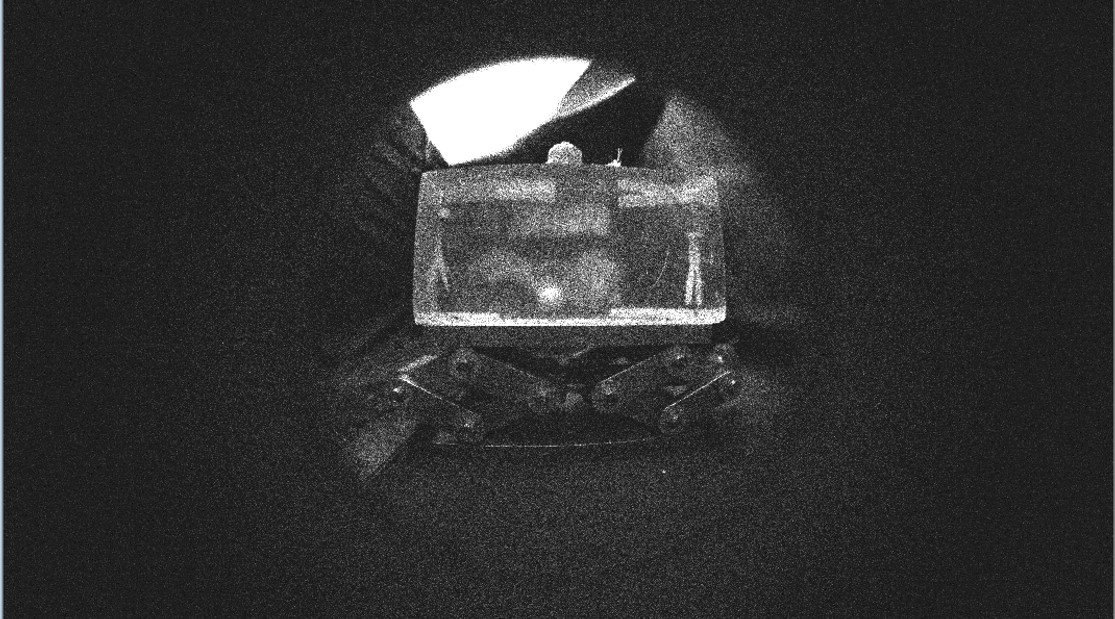}
    \caption{15\,cm}
	\label{fig:crystal:visual:15cm}       % Give a unique label
	\end{subfigure}	
    \begin{subfigure}[c]{0.45\textwidth}
	\includegraphics[width=\linewidth]{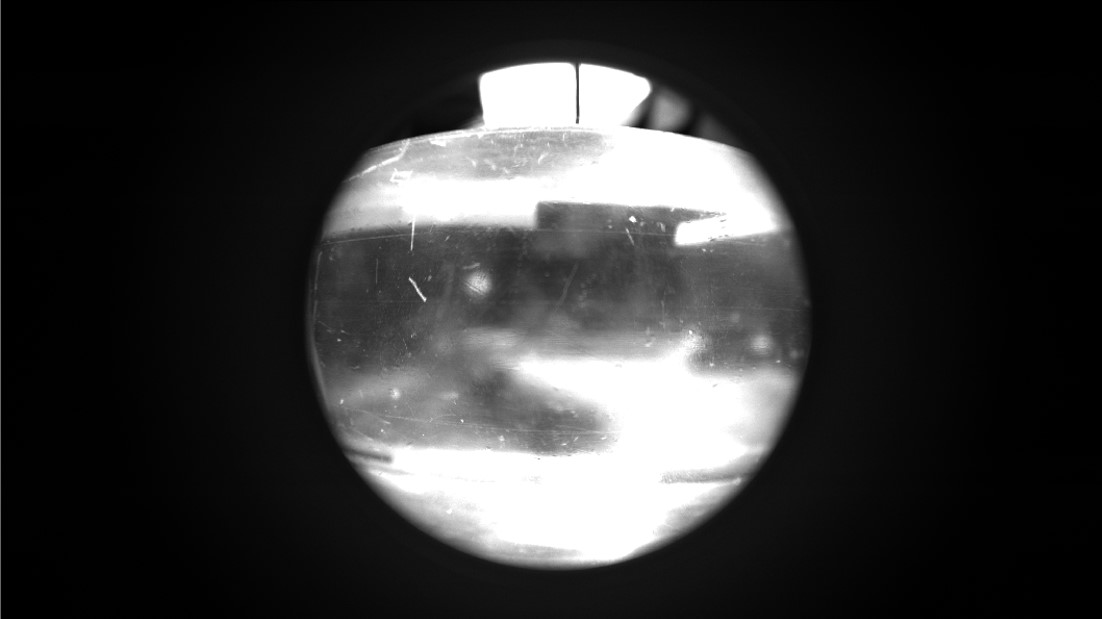}
    \caption{4\,cm}
	\label{fig:crystal:visual:4cm}       % Give a unique label
	\end{subfigure}	
    \caption{Crystal photos when the dark box is open with natural illumination and different object distances.}
    \label{fig:crystal:visual}
\end{figure*}

In order to increase the acceptance of the emitted photons from the crystal, a lens with a much short focal length and small number aperture (1/2”, C type, 6-$\infty$\,mm, f/1.4) is used. The images of the crystal with different distances are shown in Fig.\,\ref{fig:crystal:visual}, which are taken with illumination before the dark box is closed. The field of view with the used lens is in a circle and much smaller than the full size of the camera sensor. The circle shape and its outside of the field of view will be considered in the following measurement and analysis. Please note that there is a clear distortion around the edge of the field of view (crystal region) when the object distance is too small as in Fig.\,\ref{fig:crystal:visual:4cm}.

\subsection{Calibration}
\label{sec:1:system:2:calib}

\begin{figure*}[!ht]
    \centering
    \begin{subfigure}[c]{0.45\textwidth}
	\includegraphics[width=\linewidth]{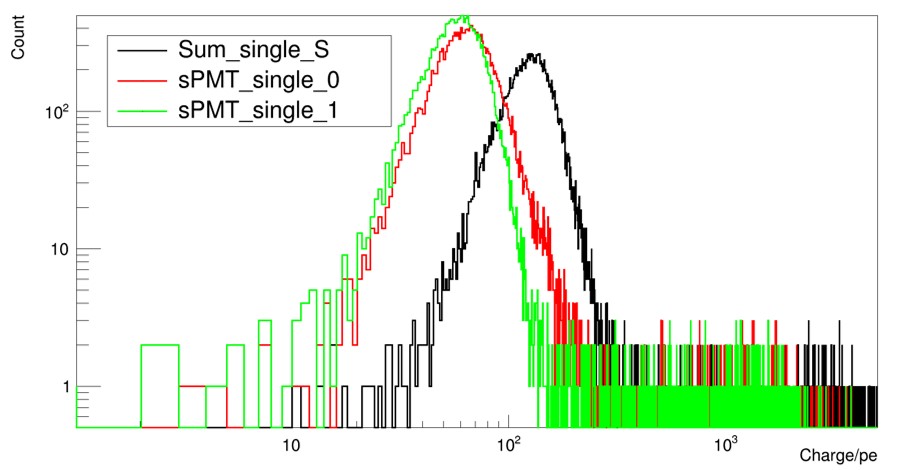}
    \caption{Charge spectra of eac PMT and sum}
	\label{fig:crystal:pmt:intensity}       % Give a unique label
	\end{subfigure}	
    \begin{subfigure}[c]{0.45\textwidth}
	\includegraphics[width=\linewidth]{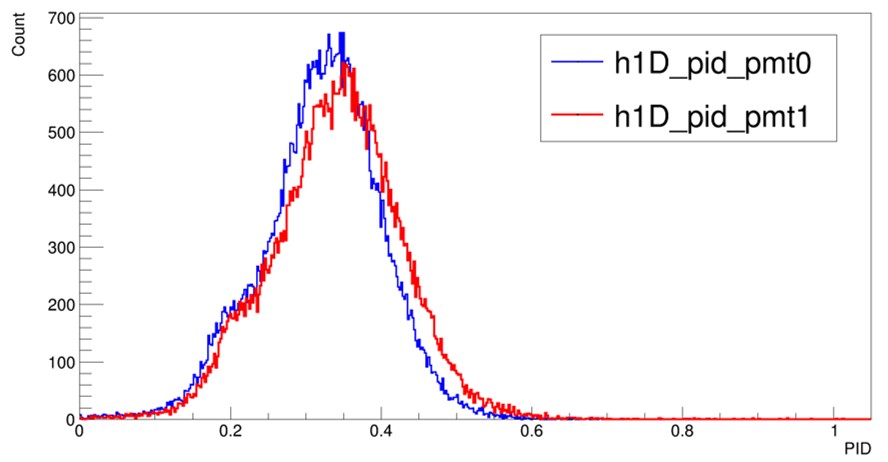}
    \caption{PID of each event of each PMT}
	\label{fig:crystal:pmt:PID}       % Give a unique label
	\end{subfigure}	
    \begin{subfigure}[c]{0.45\textwidth}
	\includegraphics[width=\linewidth]{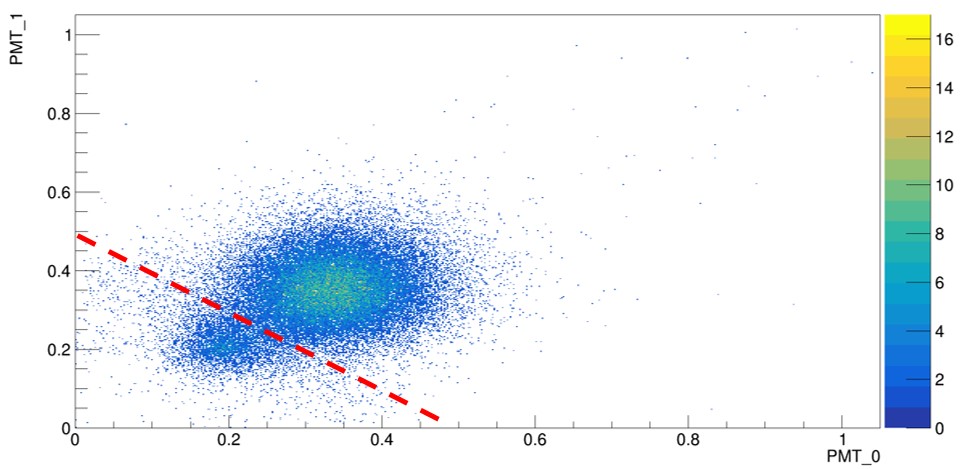}
    \caption{2-D distribution of PID of the two PMTs}
	\label{fig:crystal:pmt:PID:2D}       % Give a unique label
	\end{subfigure}	
     \begin{subfigure}[c]{0.45\textwidth}
	\includegraphics[width=\linewidth]{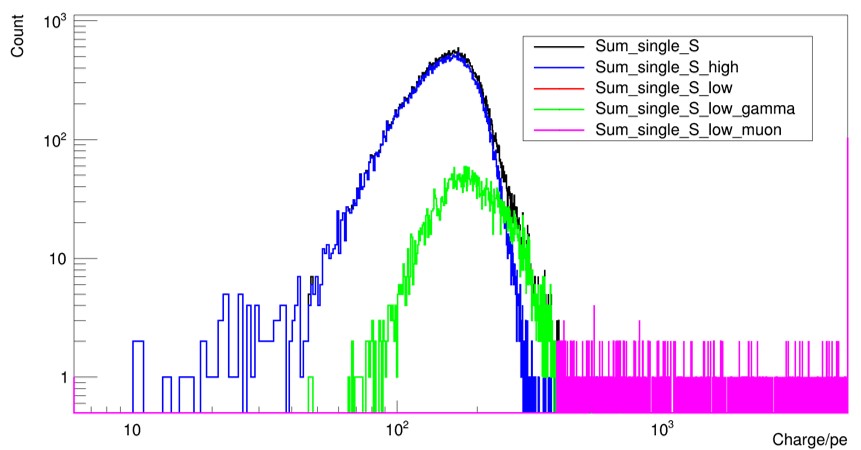}
    \caption{Selected events by PID and charge}
	\label{fig:crystal:pmt:intensity:sum}       % Give a unique label
	\end{subfigure}	
    \caption{Measured charge spectra by the 3-inch PMTs, and the PID distrbution of the waveform.}
    \label{fig:crystal:pmt}
\end{figure*}

Fig.\,\ref{fig:crystal:pmt:intensity} shows the measured charge spectra with the alpha source located when the object distance is around 15\,cm: the two 3-inch PMTs and the sum of them. A long tail can be found at the right of the spectrum, which is known as cosmic muons. A 400\,p.e.\,cut to the sum spectrum is used to select the events of muons.
A factor of particle identification (PID) is calculated by each waveform of each PMT, and it is defined by the ratio of the charge in the first 300\,ns to the whole window, as shown in Fig.\,\ref{fig:crystal:pmt:PID}. A 2-D cut on the PID is used to identify the events from the alpha source, the red dash line (PID\_pmt\_0+PID\_pmt\_1) as shown in Fig.\,\ref{fig:crystal:pmt:PID:2D}. The selected spectra are shown in Fig.\,\ref{fig:crystal:pmt:intensity:sum}, where the blue curve is selected as the alpha-like events by (PID\_pmt\_0+PID\_pmt\_1>0.5), the Magenta curve is selected for the muon events candidates by (PID\_pmt\_0+PID\_pmt\_1<0.5 and charge\_pmt\_0+charge\_pmt\_1>400\,p.e.), and the green curve is assumed as the gamma-like events by (PID\_pmt\_0+PID\_pmt\_1<0.5).

Taking into account the dead time of the data-taking system, the actual event rate of each measurement is re-normalized according to the selected muon rate, where the reference muon rate of the crystal is from the measurement and selection of without source with object distance of 15\,cm. 
It is around 3\,Hz for selected muons, 2.6\,Hz for gamma-like events, and 2.2\,Hz for alpha-like events of the measurement without source and object distance of 15\,cm. 
The data-taking rate and the re-normalized rate are listed in Table\,\ref{tab:source:rate} for different configurations. The rate of alpha-like events is around 100\,Hz and increases from 93\,Hz to 188\,Hz following the shortening of the object distance from 15\,cm to 4\,cm. The contribution from the alpha source is much higher than that from the background of around 2.2\,Hz.

Following the classification of the events, the mean charge of each kind of event is calculated too. The mean charge of the alpha-like events is 130\,p.e.\,of 15\,cm, 150\,p.e.\,of 10\,cm, and 159\,p.e.\,of 4\,cm, respectively. The signal intensity is not following the solid angle simply, for the 4\,cm, in particular, which is because the distance to the 3-inch PMTs is rather small than the object distance of the camera to the crystal according to the layout of the system. 
The mean intensity of the muon events is around 2100\,p.e., which suffers from statistic uncertainty and solid angle issues too.

\begin{table}[!htb]
\centering
\caption{Event rate and charge intensity of crystal with two 3-inch PMTs. The distance is between the camera and the crystal front surface. The events are measured by the coincidence by the two 3-inch PMTs, and the charge is from the sum of the two PMTs of each event. The alpha-like events are selected by the sum of PID, and the separation between muon and gamma-like is by a charge cut after the PID cut.}
\label{tab:source:rate}       % Give a unique label
\resizebox{\linewidth}{!}{
\begin{tabular}{c|c|c|c|c|c|c|c|c}
\hline\noalign{\smallskip}
Type  & DAQ Rate & Normalized &  \multicolumn{3}{c|}{Rate (Hz)} & \multicolumn{3}{c}{Mean Charge (p.e.)} \\
(Distance)  & (Hz)  & Rate (Hz) & Muon  & Gamma-like & Alpha-like & Muon  & Gamma-like & Alpha-like \\
\noalign{\smallskip}\hline\noalign{\smallskip}
w/o source 15\,cm  &  $\sim$8 &  7.8 &  3.0  & 2.6 & 2.2 &  2099.0 & 181.3 & 124.5 \\
\noalign{\smallskip}\cline{1-9}\noalign{\smallskip}
w/ source 15\,cm  &  $\sim$80 &  93.1 & 3.0  & 6.2  & 83.9 &  2108.9 & 184.0 & 130.2 \\
\noalign{\smallskip}\cline{1-9}\noalign{\smallskip}
w/ source 10\,cm  &  $\sim$70 &  160.1 & 3.0  & 18.2  & 138.9 &  2057.4 & 204.0 & 150.4 \\
\noalign{\smallskip}\cline{1-9}\noalign{\smallskip}
w/ source 4\,cm  &  $\sim$70 &  188.9 & 3.0  & 21.3  & 164.6 &  2062.8 & 207.1 & 159.5 \\
\noalign{\smallskip}\hline
\end{tabular}
}
\end{table}

\begin{figure*}[!ht]
    \centering
    \includegraphics[scale=0.55]{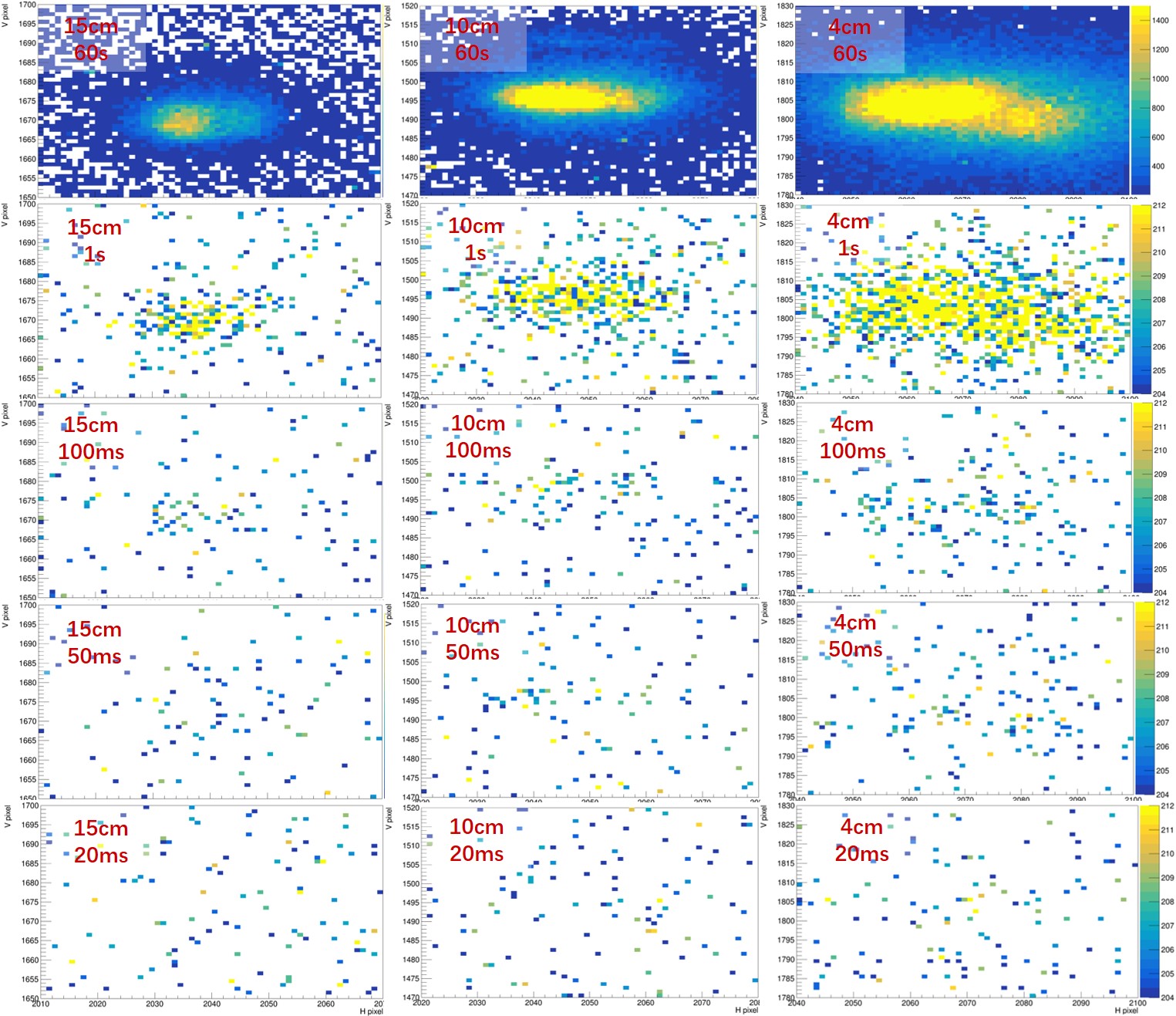}
    \caption{2-D images of alpha source with object distances of 15\,cm (50\,pixel 
 (V) $\times$60\,pixels (H), left), 10\,cm (50\,pixel (V) $\times$60\,pixels (H), middle) and 4\,cm (50\,pixel (V) $\times$60\,pixels (H), right) versus exposure time of 60\,s, 1\,s, 100\,ms, 50\,ms and 20\,ms. The pixel size is 4.6\,\textmu m $\times$ 4.6\,\textmu m. The z-axis is in the unit of ADC which is the camera output per pixel directly relative to the light intensity. The baseline of each pixel is around 200\,ADC, and the gain factor is around 7.8\,ADC/p.e.}
    \label{fig:camera:intensity:ept:2D}
\end{figure*}

The images of the crystal with the alpha source are taken by the camera with different exposure times and different object distances. The region of the source is selected and shown in Fig.\,\ref{fig:camera:intensity:ept:2D}, where the selected image dimension of the sensor is around 0.23\,mm $\times$0.28\,mm with 50\,pixel (V) $\times$60\,pixels (H) and 4.6\,\textmu m $\times$ 4.6\,\textmu m per pixel. The regime of the alpha source can be identified as around 3\,mm scale of an object distance of 15\,cm, 2\,mm scale of 10\,cm, and 1\,mm scale of 4\,cm, and the light intensity gradually dims when shortening the exposure time. It is almost identified to event level with 0.05\,s exposure time but on a higher noise background, where only a few alphas occur during the time. The dimension of the image of the source is enlarging when the object distance reduces as expected.

\begin{figure*}[!ht]
    \centering
    \includegraphics[scale=0.4]{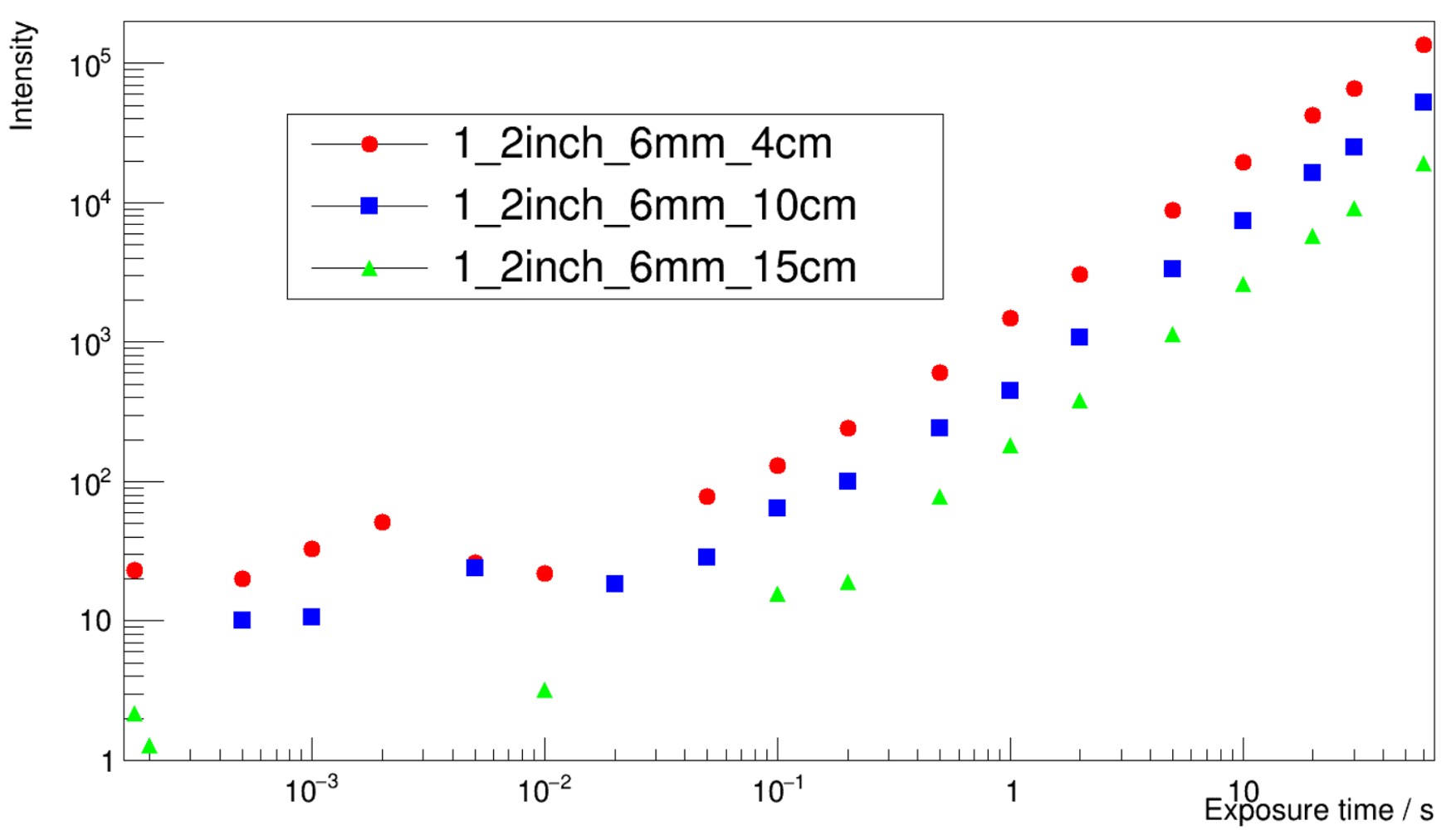}
    \caption{Measured intensity of alpha source by camera versus exposure time under different object distances.}
    \label{fig:camera:intensity:ept}
\end{figure*}

The intensity of the source region is integrated and converted into p.e.\,as shown in Fig.\,\ref{fig:camera:intensity:ept}, where the noise (baseline is around 200\,ADC) of the camera is subtracted according to a parallel region of the source with equal area\,\cite{camera-Wang:2022xgb}. The conversion factor is around 7.8\,ADC/p.e. The diameter of the source region is 21\,pixels for an object distance of 15\,cm, 22\,pixels for 10\,cm, and 33\,pixels for 4\,cm. The fitted intensity per second by a linear curve is around 314\,p.e.\,of an object distance of 15\,cm, 862\,p.e.\,of 10\,cm, and 2255\,p.e.\,of 4\,cm, respectively. 
Considering the rate of the alpha source measured under different object distances as in Tab.\,\ref{tab:source:rate}, the ratio of measured charge intensity of the camera and the PMTs is around 3\% of an object distance of 15\,cm, 4\% of 10\,cm, and 8\% of 4\,cm, respectively. 
The expected typical charge intensity of alpha-like event viewed by the camera is around 3.9\,p.e.\,of an object distance of 15\,cm, 6.0\,p.e.\,of 10\,cm, and 12.8\,p.e.\,of 4\,cm, respectively.
The expected typical charge intensity of each muon viewed by the camera is around 60\,p.e.\,of an object distance of 15\,cm, 85\,p.e.\,of 10\,cm, and 177\,p.e.\,of 4\,cm, respectively.

\section{Muon track}
\label{sec:1:muon}

\subsection{Signal vs. Noise}
\label{sec:1:muon:2:noise}

As stated in \cite{camera-Wang:2022xgb}, the noise of the camera is still much higher than the traditional used PMT or SiPM, which is much worse when we are trying to use many pixels for imaging measurement. It can be expected that it will help to identify the target by a smaller area and stronger intensity of the same object, as seen in the left plateau of the curves in Fig.\,\ref{fig:camera:intensity:ept}, where the difference of the plateau level (noise) is mainly from the dimension of the imaging area. The minimum of the plateau is from the object distance of 15\,cm configuration, which is proportional to the ratio squared of the focal length to object distance, even the final signal-to-noise (SN) ratio is also proportional to the square of the reciprocal of focal length except for the effective aperture. 

\begin{figure*}[!ht]
    \centering
    \begin{subfigure}[c]{0.45\textwidth}
	\includegraphics[width=\linewidth]{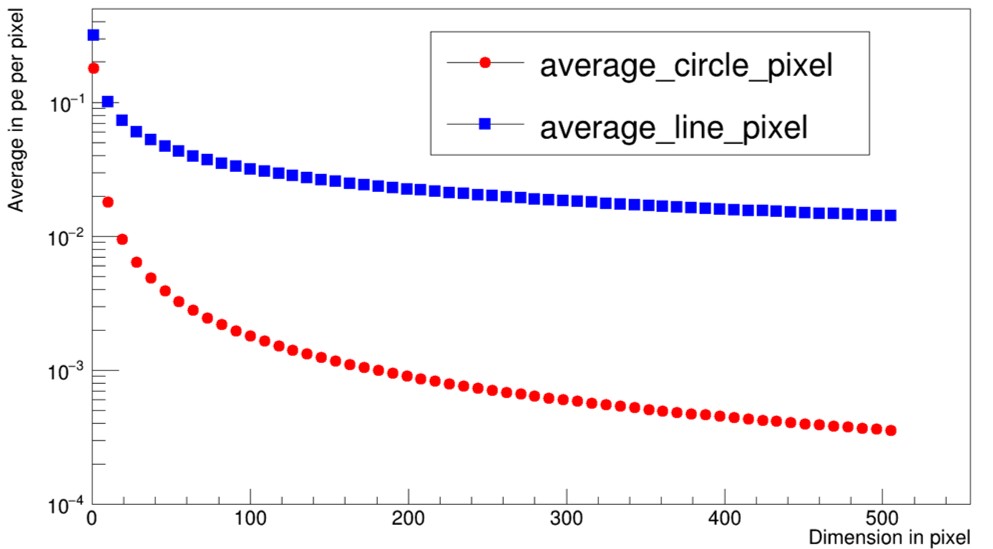}
    \caption{Uncertainty of the mean of each pixel: in p.e.}
	\label{fig:cal:average}       % Give a unique label
	\end{subfigure}	
    \begin{subfigure}[c]{0.45\textwidth}
	\includegraphics[width=\linewidth]{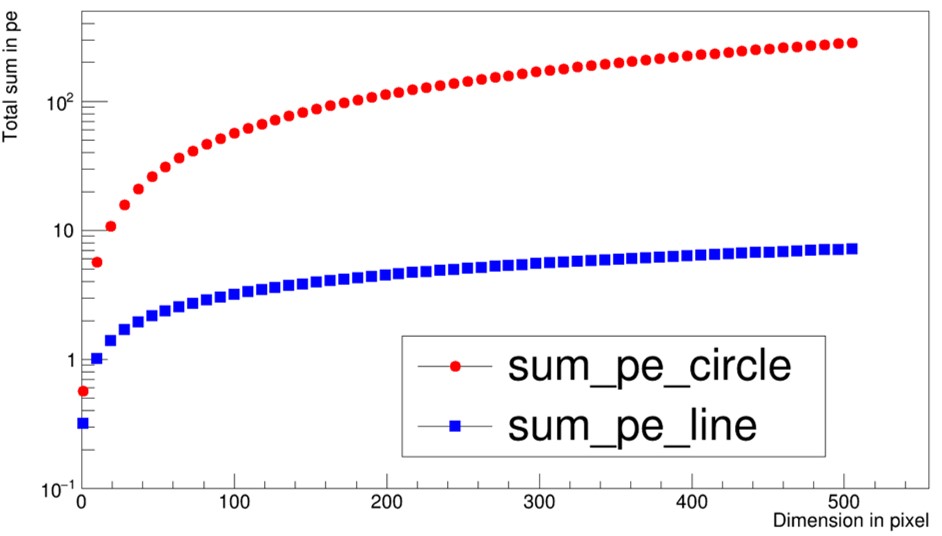}
    \caption{Sum of noise vs.\,dimension: in p.e.}
	\label{fig:cal:sum}       % Give a unique label
	\end{subfigure}	
    \caption{Pure noise versus image dimension in circles or lines.}
    \label{fig:cal:intensity}
\end{figure*}

A calculation is further executed to compare the effect of the imaging shape under the same noise level. Here a pure statistic model is used to check the mean of each pixel and total sum with an assumption of 0.3\,p.e.\,noise level per pixel (sigma of Gaussian) in a circle (diameter) or line (in length and in width of one pixel) shape. The uncertainty of the mean of each pixel is inversely proportional to the square root of total pixel numbers as shown in Fig.\,\ref{fig:cal:average}, where more pixels will have smaller fluctuation comparing circles to lines. While the total sum of all pixels is proportional to the square root of pixel numbers shown in Fig.\,\ref{fig:cal:sum}, where the required sum in lines is much smaller than that of in circles to identify a signal. If we aim to identify a signal by a three-times signal-to-noise ratio with similar total intensity, the signal in a line is much more effective than that in a circle.

\begin{figure*}[!ht]
    \centering
    \begin{subfigure}[c]{0.45\textwidth}
	\includegraphics[width=\linewidth]{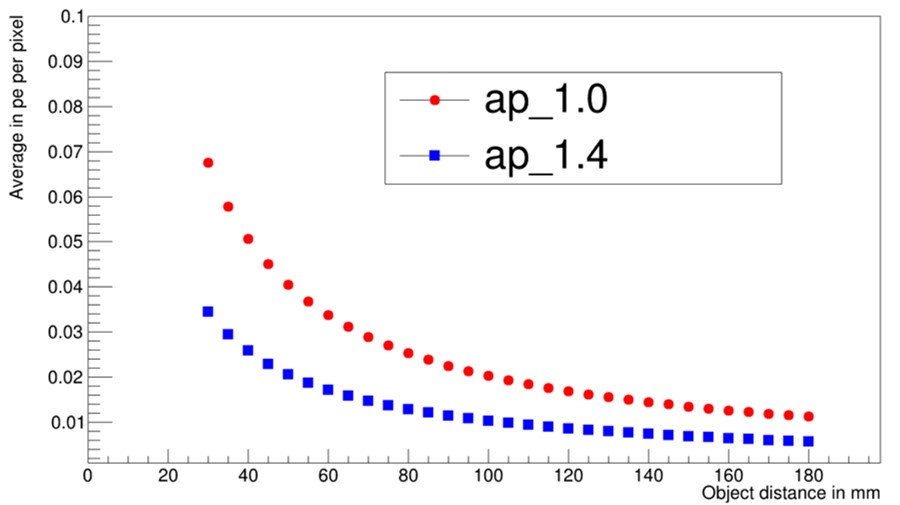}
    \caption{Averaged intensity of muon track in pixel}
	\label{fig:cal:pixel}       % Give a unique label
	\end{subfigure}	
    \begin{subfigure}[c]{0.45\textwidth}
	\includegraphics[width=\linewidth]{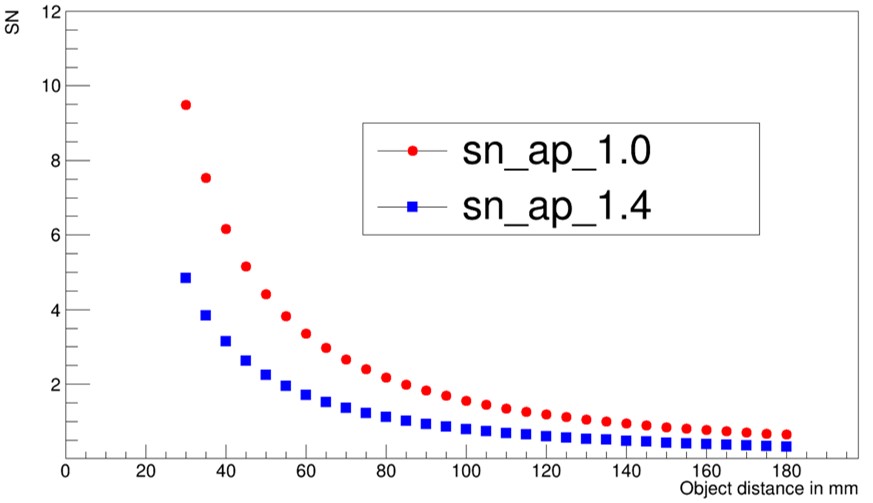}
    \caption{S/N expectation following the model}
	\label{fig:cal:sn0}       % Give a unique label
	\end{subfigure}	
    \caption{Intensity and signal-to-noise ratio of a muon track: 20000\,photons/MeV, 2\,MeV/cm, focal length 6\,mm, sensor quantum efficiency 30\%, a 4\,cm track, around 8\,p.e./cm and 330\,pixels/cm at object distance of 4\,cm.}
    \label{fig:cal:sn}
\end{figure*}

Following this strategy, an alpha-like event in a circle is difficult to be identified directly according to their aimed imaging area and captured light intensity if we only can locate a range of an image by one camera as in \cite{camera-Wang:2022xgb}.
While, in another hand, a muon track is a good candidate to check if we assume its image follows a straight line with tiny width.
With the configurations of the crystal system (20000\,photons/MeV, 2\,MeV/cm, focal length 6\,mm, sensor quantum efficiency 30\%, a 4\,cm track, around 8\,p.e./cm and 330\,pixels/cm at an object distance of 4\,cm), we can anticipate the averaged intensity of each pixel along the muon track imaging versus object distance and effective numerical aperture (ap), as shown in Fig.\,\ref{fig:cal:pixel}. A smaller object distance and effective numerical aperture mean more light is collected with the same focal length and lens diameter. The expected signal-to-noise ratio with the assumption of 0.3\,p.e.\,noise, and 4\,cm track length can be further checked as shown in Fig.\,\ref{fig:cal:sn0}, and it is around three times SN when the object distance is 4\,cm.

\begin{figure*}[!ht]
    \centering
    \begin{subfigure}[c]{0.45\textwidth}
	\includegraphics[width=\linewidth]{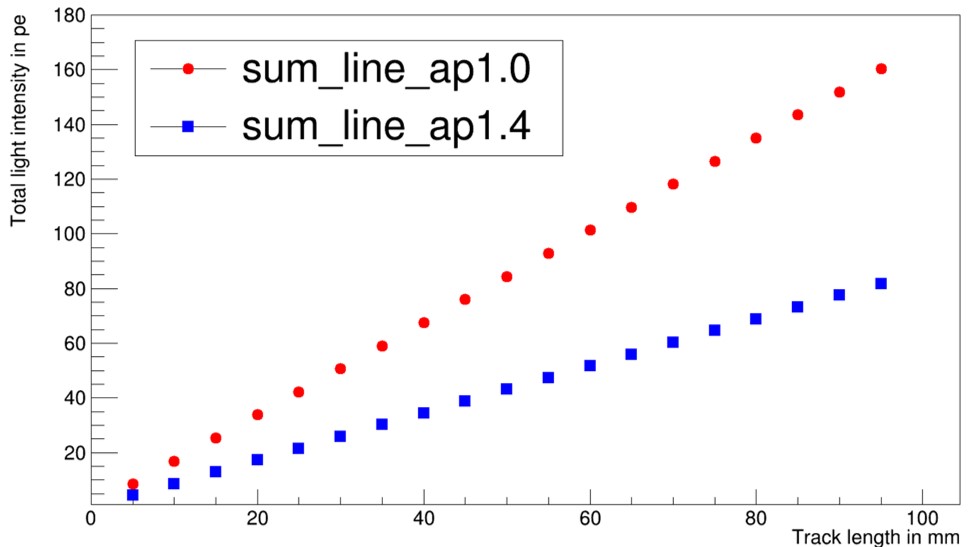}
    \caption{Total intensity versus track length}
	\label{fig:cal:tracklength:sum}       % Give a unique label
	\end{subfigure}	
    \begin{subfigure}[c]{0.45\textwidth}
	\includegraphics[width=\linewidth]{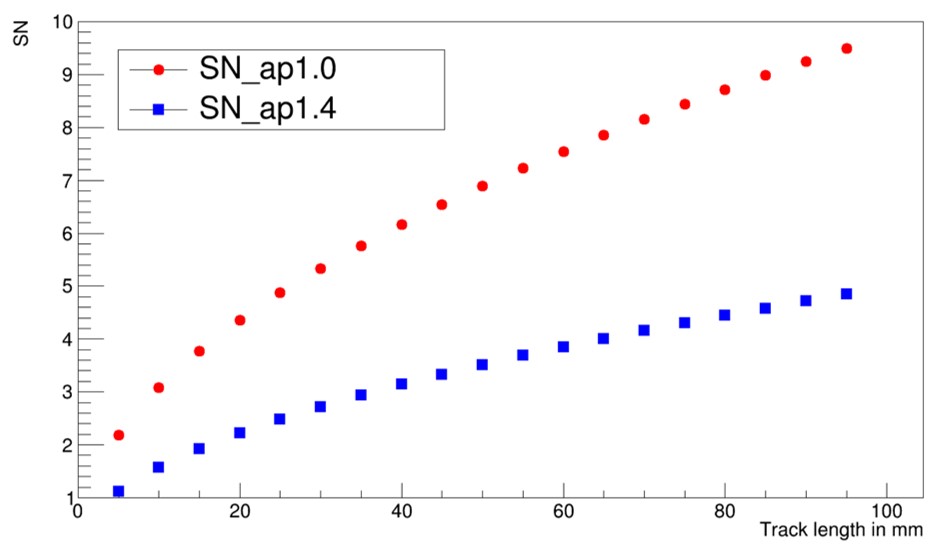}
    \caption{S/N versus track length}
	\label{fig:cal:tracklength:sn}       % Give a unique label
	\end{subfigure}	
     \begin{subfigure}[c]{0.45\textwidth}
	\includegraphics[width=\linewidth]{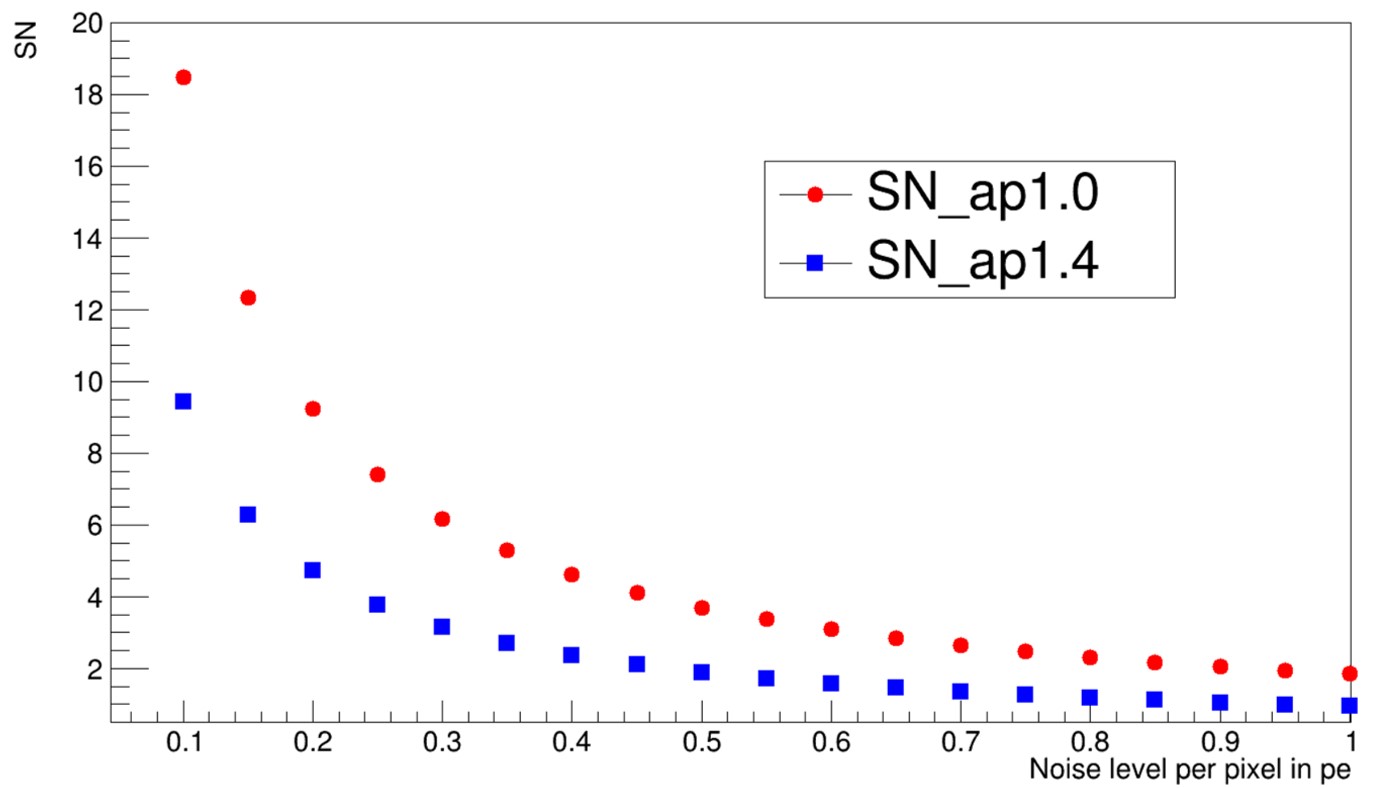}
    \caption{S/N versus noise level per pixel}
	\label{fig:cal:noise:sn}       % Give a unique label
	\end{subfigure}	
    \caption{Total intensity and signal-to-noise ratio versus track length, and signal-to-noise ratio versus noise level per pixel and effective numerical aperture with a 4\,cm track: 20000\,photons/MeV, 2\,MeV/cm, focal length 6\,mm, sensor quantum efficiency 30\%, around 8\,p.e./cm and 330\,pixels/cm at an object distance of 4\,cm.}
    \label{fig:cal:noise:sn:tracklength}
\end{figure*}

The measured total intensity of a track will increase following the track length as shown in Fig.\,\ref{fig:cal:tracklength:sum}, where a length of 10\,mm track means around 330\,pixels with 6\,mm focal length and 4\,cm object distance. The signal-to-noise ratio also will improve following the track length increasing as in Fig.\,\ref{fig:cal:tracklength:sn}.
The effect of the noise level in each pixel is further evaluated in Fig.\,\ref{fig:cal:noise:sn}. It will reach around three times the signal-to-noise level with 0.3\,p.e.\,noise level, 4\,cm track length, 6\,mm focal length, 1.4 aperture, and 4\,cm object distance. A smaller noise per pixel and smaller aperture will help improve the signal-to-noise ratio as well as larger optical lens dimensions.

\subsection{Image track survey}
\label{sec:1:muon:2:track}

\begin{figure*}[!ht]
    \centering
    \begin{subfigure}[c]{0.45\textwidth}
	\includegraphics[width=\linewidth]{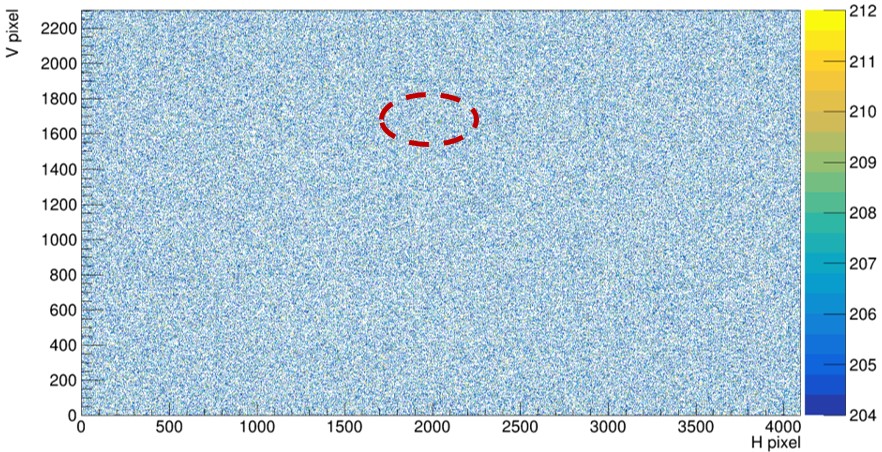}
    \caption{Object distance of 15\,cm}
	\label{fig:camera:image:full:2D:15cm}       % Give a unique label
	\end{subfigure}	
    \begin{subfigure}[c]{0.45\textwidth}
	\includegraphics[width=\linewidth]{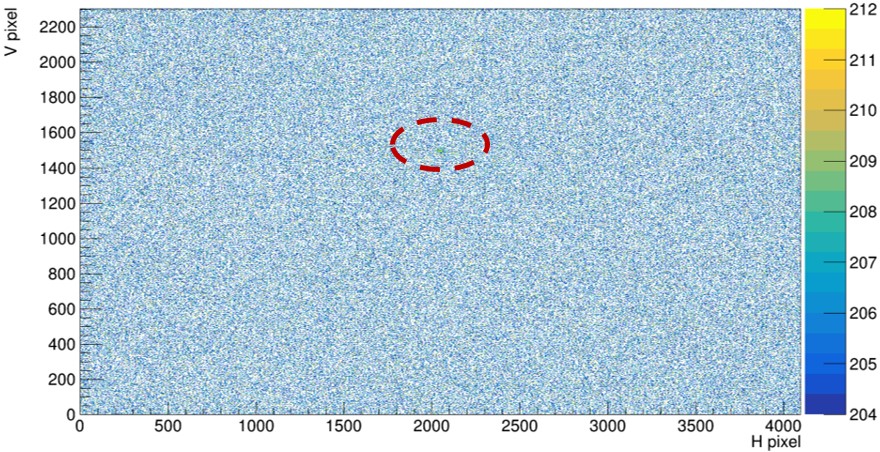}
    \caption{Object distance of 10\,cm}
	\label{fig:camera:image:full:2D:10cm}       % Give a unique label
	\end{subfigure}	
     \begin{subfigure}[c]{0.45\textwidth}
	\includegraphics[width=\linewidth]{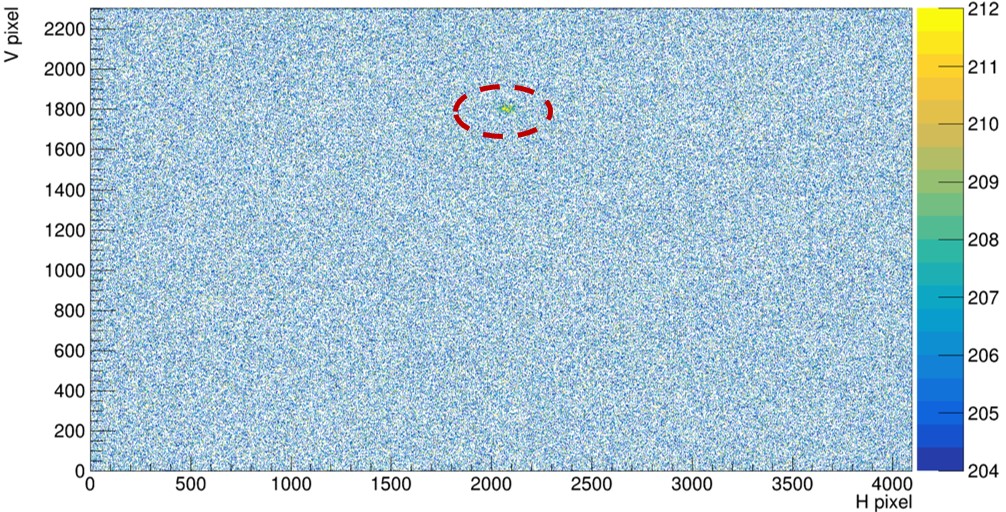}
    \caption{Object distance of 4\,cm}
	\label{fig:camera:image:full:2D:4cm}       % Give a unique label
	\end{subfigure}	
    \caption{Full image of the crystal with alpha source under 1\,s exposure time.}
    \label{fig:camera:image:full:2D}
\end{figure*}

The images of the crystal system with the alpha source and 1\,s exposure time are taken.
Fig.\,\ref{fig:camera:image:full:2D} shows an example of the image, where the location of the alpha source can be identified clearly, and it will be used as an online anchor for light intensity and location.
While according to the calculation in Sec.\,\ref{sec:1:muon:2:noise}, the object distance of 4\,cm configuration is run continuously for 30\,s to check out possible muon racks with better signal-to-noise ratio.

\begin{figure*}[!ht]
    \centering
    \begin{subfigure}[c]{0.45\textwidth}
	\includegraphics[width=\linewidth]{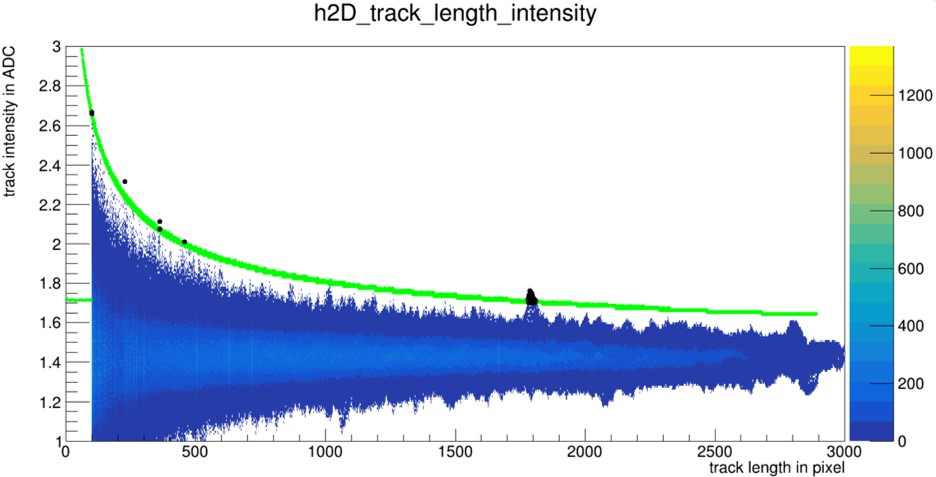}
    \caption{Averaged intensity of each pixel 2-D}
	\label{fig:crystal:survey:avg:2D}       % Give a unique label
	\end{subfigure}	
     \begin{subfigure}[c]{0.45\textwidth}
	\includegraphics[width=\linewidth]{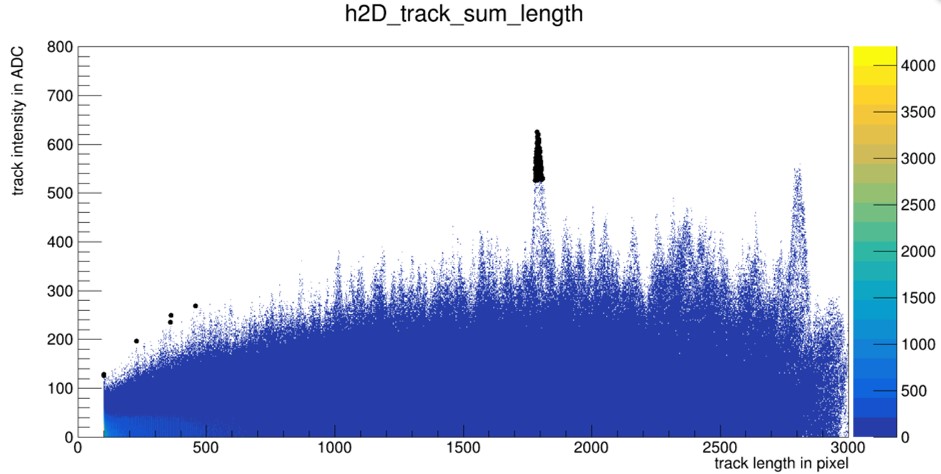}
    \caption{Intensity sum 2-D}
	\label{fig:crystal:survey:sum:2D}       % Give a unique label
	\end{subfigure}	
    \begin{subfigure}[c]{0.45\textwidth}
	\includegraphics[width=\linewidth]{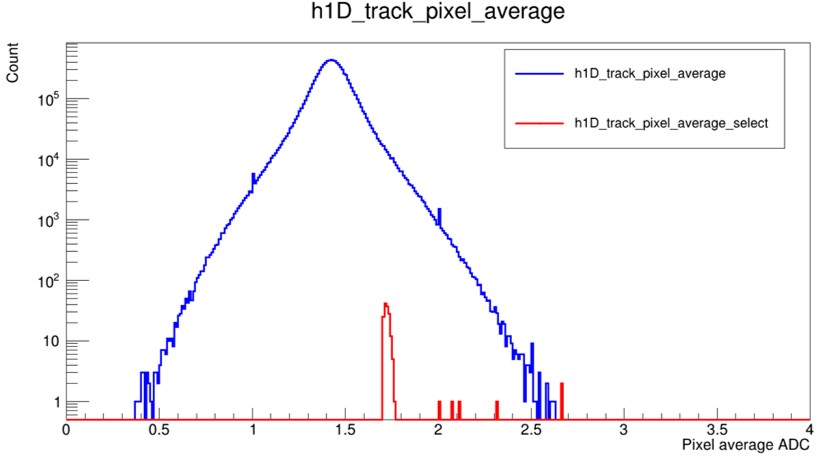}
    \caption{Averaged intensity of each pixel 1-D}
	\label{fig:crystal:survey:avg:1D}       % Give a unique label
	\end{subfigure}	
    \begin{subfigure}[c]{0.45\textwidth}
	\includegraphics[width=\linewidth]{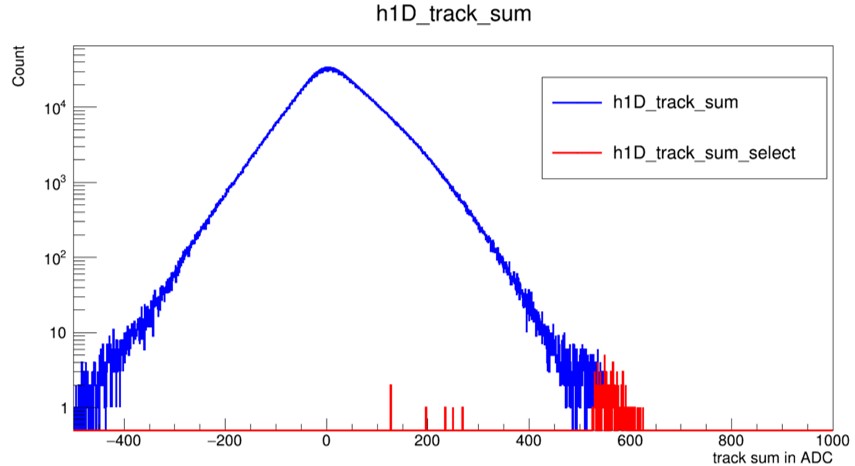}
    \caption{Intensity sum 1-D}
	\label{fig:crystal:survey:sum:1D}       % Give a unique label
	\end{subfigure}	
    \caption{Averaged intensity of each pixel and the sum of the assumed tracks.}
    \label{fig:crystal:survey}
\end{figure*}

With the images (object distance of 4\,cm as an example), a survey was done to each assumed line in aimed pixel range (from vertical line 1700 to vertical line 1600, the minimum track length is around 100\,pixels) following the strategy, where the averaged intensity of each pixel is calculated among each assumed lines as shown in Fig.\,\ref{fig:crystal:survey:avg:2D} and Fig.\,\ref{fig:crystal:survey:avg:1D}. In Fig.\,\ref{fig:crystal:survey:avg:2D}, a green dotted line is also plotted, which is a used cut of five times of noise uncertainty which is relative to the pixel number in a model (0.3*7.8\,ADC/sqrt(pixel number)). Some tracks can be identified as tagged in red and shown in Fig.\,\ref{fig:crystal:survey:avg:1D}. The sum of each assumed track is also plotted in Fig.\,\ref{fig:crystal:survey:sum:2D} and Fig.\,\ref{fig:crystal:survey:sum:1D} including the selected possible track candidates. An offset of the pixel average is related to the baseline of each pixel which is further corrected during the sum calculation. The five-times cut selects the possible tracks efficiently from noise, which is better for long track and higher than the calculated three-times to avoid more noise as seen.

\begin{figure*}[!ht]
    \centering
    \includegraphics[scale=0.7]{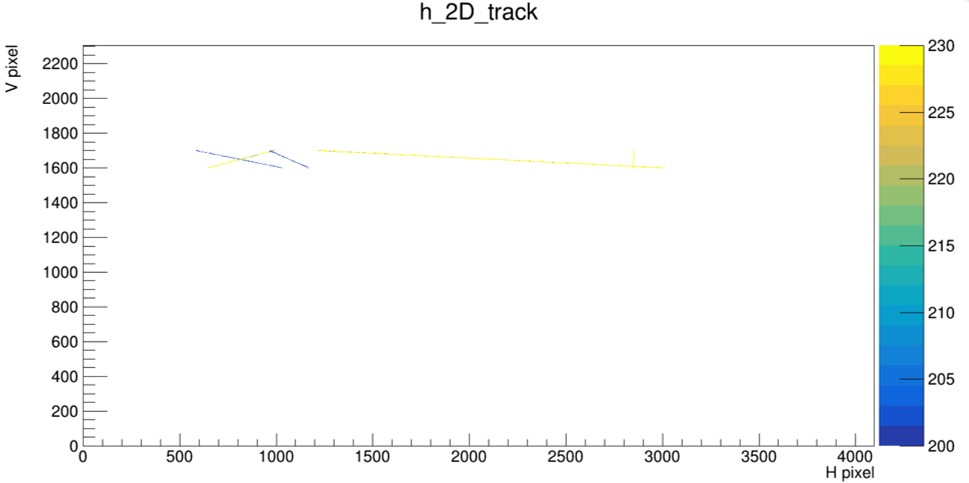}
    \caption{Selected tracks under 1\,s exposure time.}
    \label{fig:camera:track}
\end{figure*}

The identified candidate of tracks by the survey can be found in Fig.\,\ref{fig:camera:track}, while it seems too many than the expectation which is around three muon per second hitting the crystal. The direction of the candidate tracks and locations also exceed the range of the view field of the lens in Fig.\,\ref{fig:crystal:visual}. It still needs further checking on the quality of the identification even if a five-times cut is used. 

\subsection{Muon identification}
\label{sec:1:muon:2:track:ID}

\begin{figure*}[!ht]
    \centering
    \begin{subfigure}[c]{0.32\textwidth}
	\includegraphics[width=\linewidth]{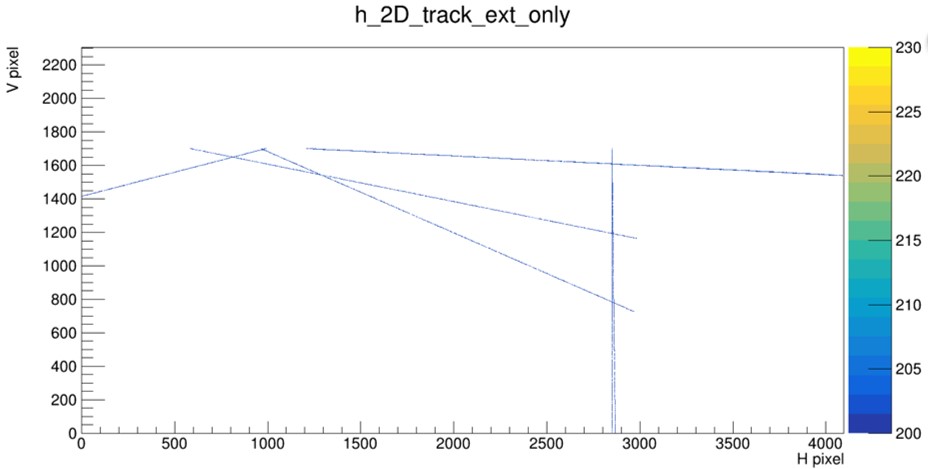}
    \caption{Extended tracks}
	\label{fig:crystal:ID:confirm:track}       % Give a unique label
	\end{subfigure}	
    \begin{subfigure}[c]{0.32\textwidth}
	\includegraphics[width=\linewidth]{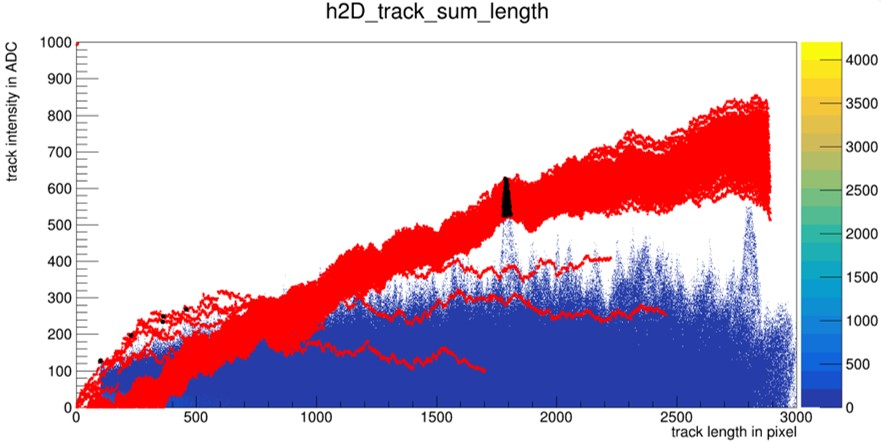}
    \caption{Extended sum}
	\label{fig:crystal:ID:confirm:sum:2D}       % Give a unique label
	\end{subfigure}	
    \begin{subfigure}[c]{0.32\textwidth}
	\includegraphics[width=\linewidth]{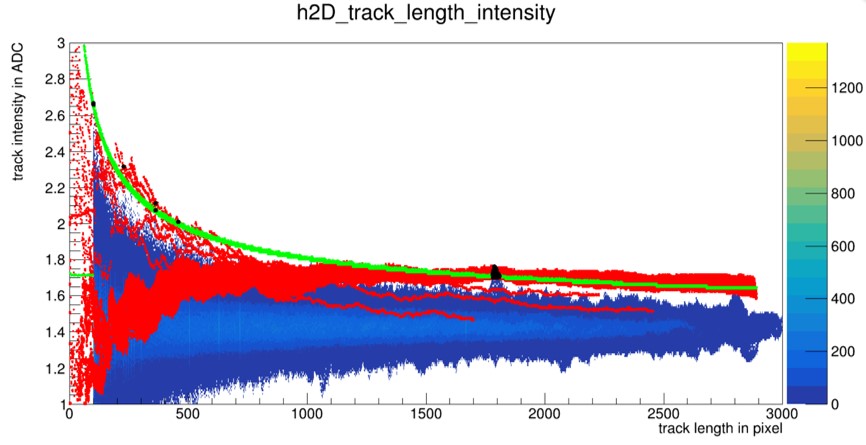}
    \caption{Extended average}
	\label{fig:crystal:ID:confirm:track:avg:2D}       % Give a unique label
	\end{subfigure}	
    \caption{Track candidate checking by extension.}
    \label{fig:crystal:ID:confirm}
\end{figure*}

The identified track candidates need to be further checked as muon track candidates. Checking the uniform trend along the tracking candidate is a good way to avoid the noise effect. It is extended to another 2000 pixels along the track in maximum (Fig.\,\ref{fig:crystal:ID:confirm:track}) to check the summed intensity (Fig.\,\ref{fig:crystal:ID:confirm:sum:2D}) and its averaged intensity per pixel versus the track's length (Fig.\,\ref{fig:crystal:ID:confirm:track:avg:2D}). As shown in Fig.\,\ref{fig:crystal:ID:confirm:sum:2D}, most of the candidates are excluded after the extension, and the sum of a few candidates keeps increasing versus the length of a true track candidate as expected in Fig.\,\ref{fig:cal:tracklength:sum}. As shown in Fig.\,\ref{fig:crystal:ID:confirm:track:avg:2D}, most of the candidates are excluded too, and the average of a few candidates keeps increasing or stable versus the length which is over the cut.

\begin{figure*}[!ht]
    \centering
    \begin{subfigure}[c]{0.45\textwidth}
	\includegraphics[width=\linewidth]{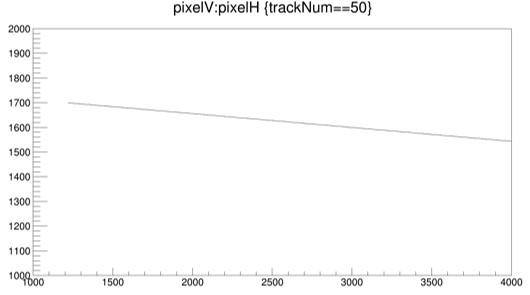}
    \caption{Extended track 1-D}
	\label{fig:crystal:track:example:1D}       % Give a unique label
	\end{subfigure}	
    \begin{subfigure}[c]{0.45\textwidth}
	\includegraphics[width=\linewidth]{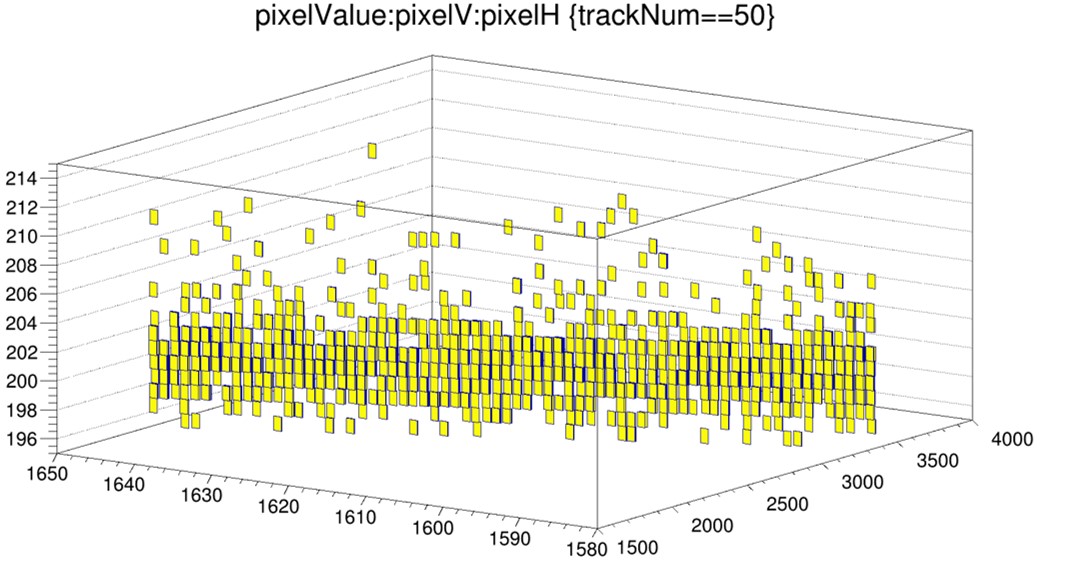}
    \caption{Extended track in 3-D}
	\label{fig:crystal:track:example:3D:2D}       % Give a unique label
	\end{subfigure}	
    \caption{Track candidates one in 3-D}
    \label{fig:crystal:track:example}
\end{figure*}

\begin{figure*}[!ht]
    \centering
    \begin{subfigure}[c]{0.45\textwidth}
	\includegraphics[width=\linewidth]{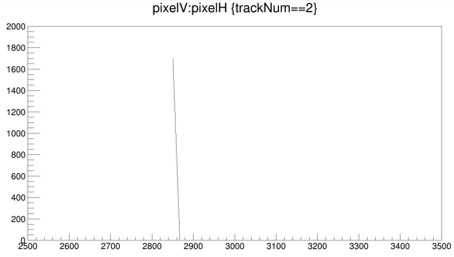}
    \caption{Extended track: 1-D}
	\label{fig:crystal:track:example:1D2}       % Give a unique label
	\end{subfigure}	
    \begin{subfigure}[c]{0.45\textwidth}
	\includegraphics[width=\linewidth]{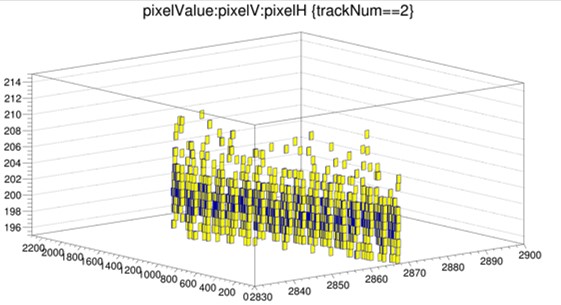}
    \caption{extended track: 3-D}
	\label{fig:crystal:track:example:3D2}       % Give a unique label
	\end{subfigure}	
    %\begin{subfigure}[c]{0.3\textwidth}
	%\includegraphics[width=\linewidth]{figures/crystal_alphasource_num_51_track_2_3D_2D.jpg}
    %\caption{fig:crystal:track:example}
	%\label{fig:crystal:track:example:3D:2D2}       % Give a unique label
	%\end{subfigure}	
    \caption{Track candidates two in 3-D}
    \label{fig:crystal:track:example2}
\end{figure*}

The tracks after further checking are drawn in the 3-D plots as shown in Fig.\,\ref{fig:crystal:track:example} and Fig.\,\ref{fig:crystal:track:example2}. Fig.\,\ref{fig:crystal:track:example} shows a candidate with a short length, the direction (Fig.\,\ref{fig:crystal:track:example:1D}), and intensity (Fig.\,\ref{fig:crystal:track:example:3D:2D}) distribution is reasonable. Fig.\,\ref{fig:crystal:track:example2} shows a candidate with a long length, the direction (Fig.\,\ref{fig:crystal:track:example:1D2}) and intensity (Fig.\,\ref{fig:crystal:track:example:3D2}) distribution is reasonable too.  They are good candidates for muon tracks. 

\section{Discussion}
\label{sec:1:dis}

\subsection{Muon tracks or not?}
\label{sec:1:dis:2:dis}

\begin{figure*}[!ht]
    \centering
    \begin{subfigure}[c]{0.45\textwidth}
	\includegraphics[width=\linewidth]{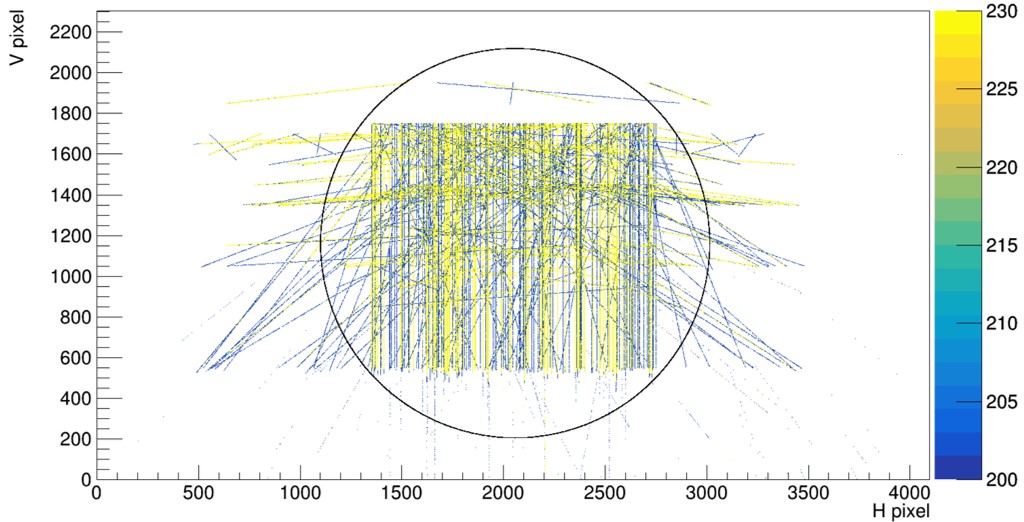}
    \caption{All selected tracks}
	\label{fig:crystal:tracks}       % Give a unique label
	\end{subfigure}	
    \begin{subfigure}[c]{0.45\textwidth}
	\includegraphics[width=\linewidth]{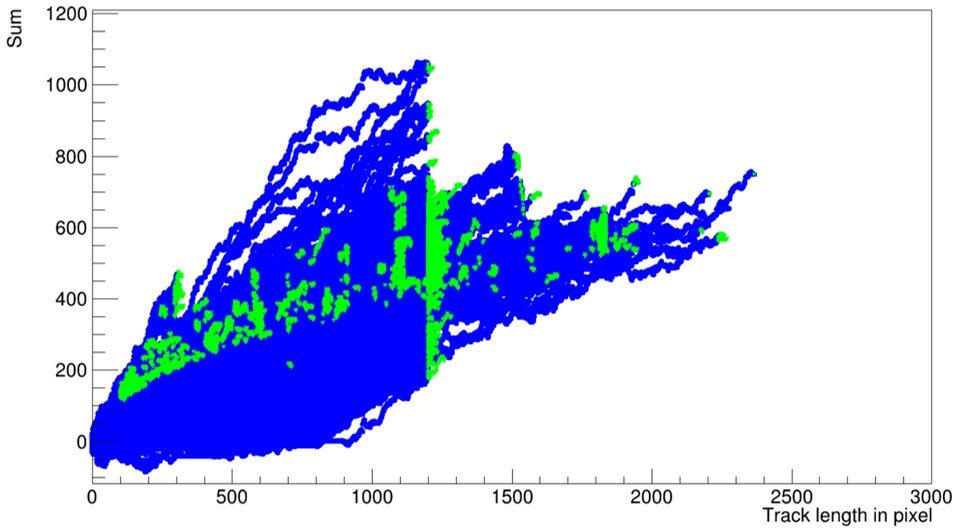}
    \caption{Total intensity in ADC vs.\, track length}
	\label{fig:crystal:track:sum2D}       % Give a unique label
	\end{subfigure}	
    \begin{subfigure}[c]{0.45\textwidth}
	\includegraphics[width=\linewidth]{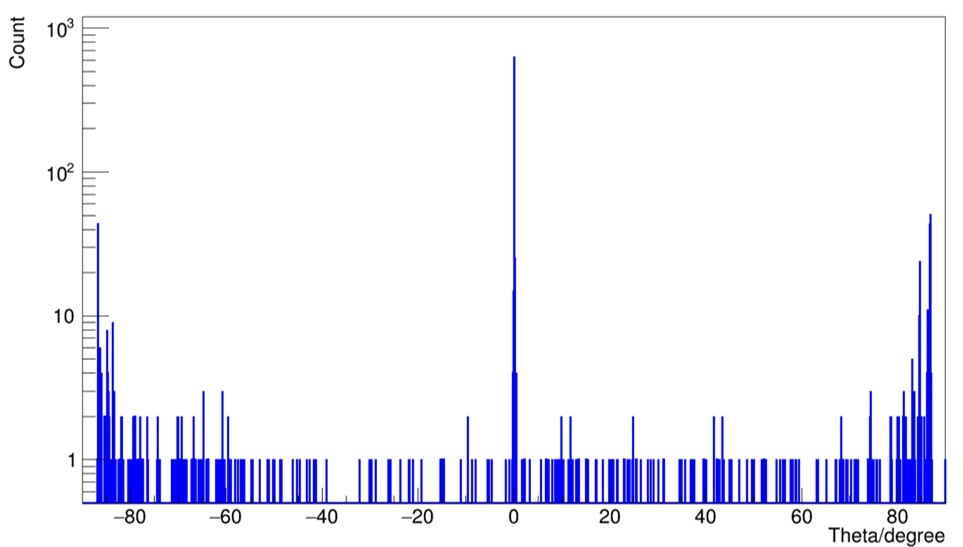}
    \caption{Theta in degree}
	\label{fig:crystal:track:angle:theta:all}       % Give a unique label
	\end{subfigure}	
    \begin{subfigure}[c]{0.45\textwidth}
	\includegraphics[width=\linewidth]{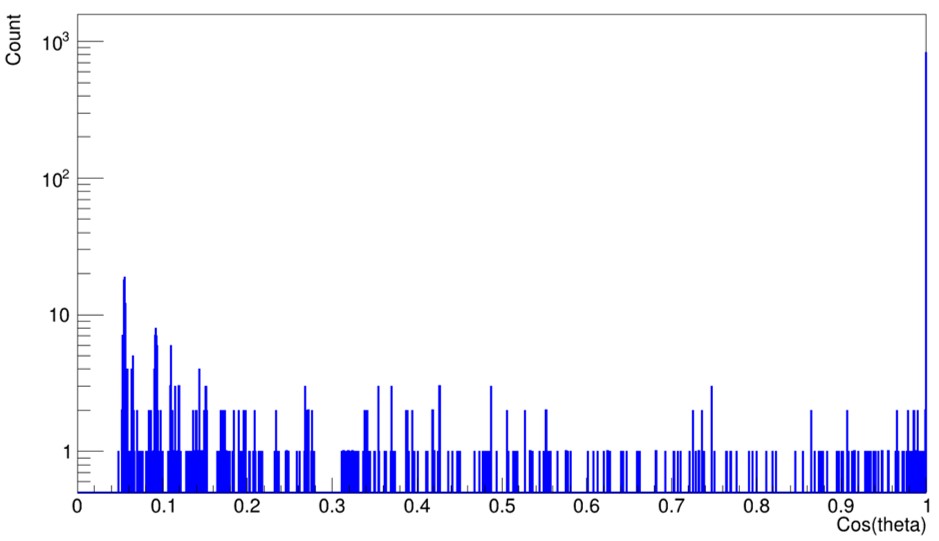}
    \caption{Cos(theta)}
	\label{fig:crystal:track:angle:costheta}       % Give a unique label
	\end{subfigure}	
    \caption{Distribution of selected tracks}
    \label{fig:camera:track:thetqa2}
\end{figure*}

The angle distribution of the selected muon track candidates is further plotted as in Fig.\,\ref{fig:camera:track:thetqa2} after the track extension check as shown in Fig.\,\ref{fig:crystal:track:sum2D}, where the theta angle is defined in the range [-90,+90]$^{\circ}$ to distinguish left and right relative to the camera. There are still too many abnormal tracks around 0 or |90|$^{\circ}$ as seen in Fig.\,\ref{fig:crystal:tracks} and Fig.\,\ref{fig:crystal:track:angle:theta:all}, which are related to some kind of systematic readout noise of the camera. 
After excluding the abnormal tracks around 0 or |90|$^{\circ}$, the theta distribution of the track candidates is basically consistent with expectation and a peak around 40$^{\circ}$ (cos($\theta$)$\sim$0.7) as a hint (Fig.\,\ref{fig:crystal:track:angle:theta:all} and Fig.\,\ref{fig:crystal:track:angle:costheta}). But the statistics are still not enough to have a good check by the muon angle distribution, even the selected muon candidates are much more than the expectation during the 30\,s data-taking period (around 3\,Hz $\times$ 30\,s).

According to \cite{am241-PhysRev.87.277,TRETYAK201040-quenching2009}, the quenching factor of the alpha of $^{241}$Am in CsI(Tl) is taken as 0.5 with an energy of 5.4\,MeV, then the light intensity viewed by the camera is around 1.4\,p.e./MeV at an object distance of 15\,cm, 2.2\,p.e./MeV of 10\,cm, and 4.7\,p.e./MeV of 4\,cm following the measurement in Sec.\,\ref{sec:1:system:2:calib}, respectively. Assuming the energy deposit of muon is around 2\,MeV per cm, 
it means 2.8\,p.e./400\textmu m at an object distance of 15\,cm (around 80\,pixels, 0.035\,p.e./pixel) on camera, 4.4\,p.e./625\textmu m at an object distance of 10\,cm (around 105\,pixels, 0.042\,p.e./pixel) on camera, and 9.4\,p.e./1300\textmu m at an object distance of 4\,cm (around 260\,pixels, 0.036\,p.e./pixel) on camera. With the data in Fig.\ref{fig:crystal:track:sum2D}, the intensity per pixel of the selected muon track candidates is from 0.1 to 1 ADC or 0.01 to 0.1 p.e., which is wide than the expectation from the measurement in Sec.\,\ref{sec:1:system:2:calib} and could be related the variation of muon location.

\subsection{Possible system optimization}
\label{sec:1:dis:2:update}

Taking more data to accumulate the muon tracks is an effective solution for more precise angle checking to find out more features of the noise, but the data volume will increase too. 
Following the issues of muon track identification, shorter exposure time is one of the effective methods to reduce the camera noise as known, while the data volume will increase too for similar statistics.

To realize a good muon tagging and reconstruction method by visual photons directly, except to find a camera with a further lower noise level as the skipper CCD\cite{Skipper-CCD-osti_1839558}, it is possible to further improve the system by updating crystals with higher light yield, and better apertures, and to reduce the distortion of the image by the lens.

Improving the track identification algorithms including distortion identification is also a valuable solution for track identification and better data compression.
The coincidence with more than one camera is also another effective way to reduce the noise and improve the measurement as suggested in \cite{camera-Wang:2022xgb}.

\subsection{Further applications}
\label{sec:1:dis:2:app}

\begin{figure*}[!ht]
    \centering
	\includegraphics[width=0.75\linewidth]{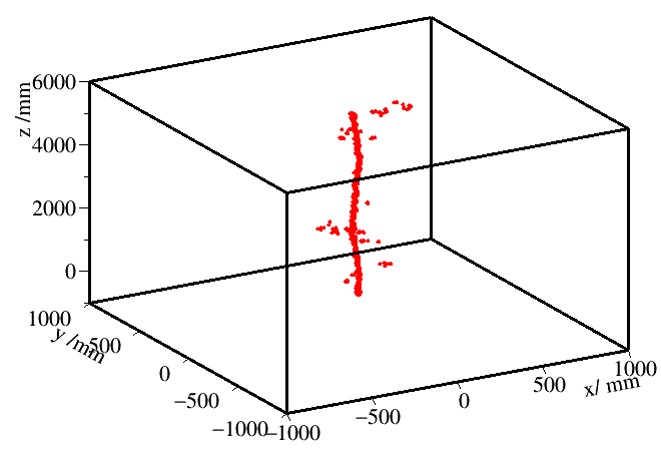}
    \caption{Simulated 1\,GeV Muon in liquid scintilator}
    \label{fig:sim:muon}
\end{figure*}

With the novel method to measure the muon track or realize a similar topology of a particle measurement in a scintillation detector, it can be used in a huge LS detector to measure the track of muons precisely as shown in Fig.\,\ref{fig:sim:muon}, including to identify possible showers along the tracks as in JUNO, which is valuable to further study relevant background and suppress their contribution to neutrino measurement.

\begin{figure*}[!ht]
    \centering
    \begin{subfigure}[c]{0.45\textwidth}
	\includegraphics[width=\linewidth]{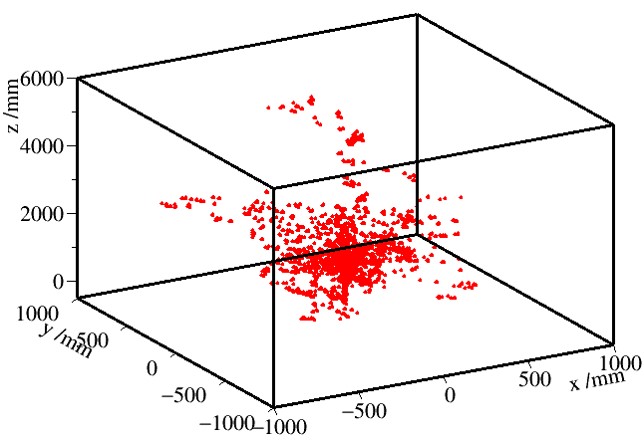}
    \caption{1\,GeV Gamma}
	\label{fig:sim:particle:electron}       % Give a unique label
	\end{subfigure}	
     \begin{subfigure}[c]{0.45\textwidth}
	\includegraphics[width=\linewidth]{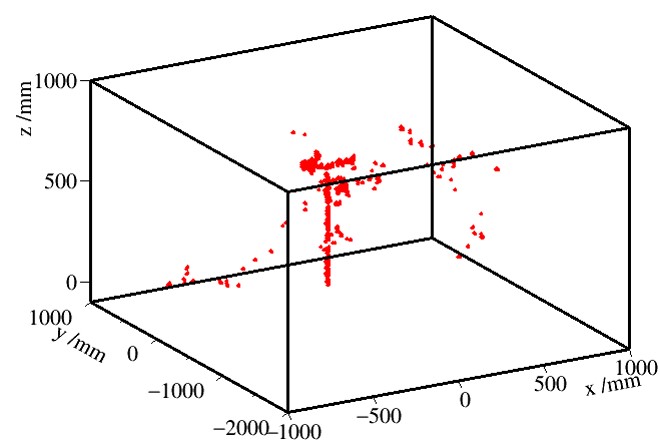}
    \caption{1\,GeV proton}
	\label{fig:sim:particle:proton}       % Give a unique label
	\end{subfigure}	
     \begin{subfigure}[c]{0.45\textwidth}
	\includegraphics[width=\linewidth]{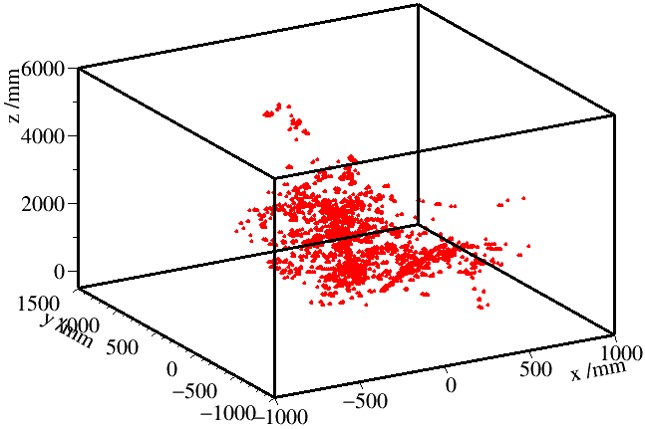}
    \caption{1\,GeV $\pi ^0$}
	\label{fig:sim:particle:pion0}       % Give a unique label
	\end{subfigure}	
     \begin{subfigure}[c]{0.45\textwidth}
	\includegraphics[width=\linewidth]{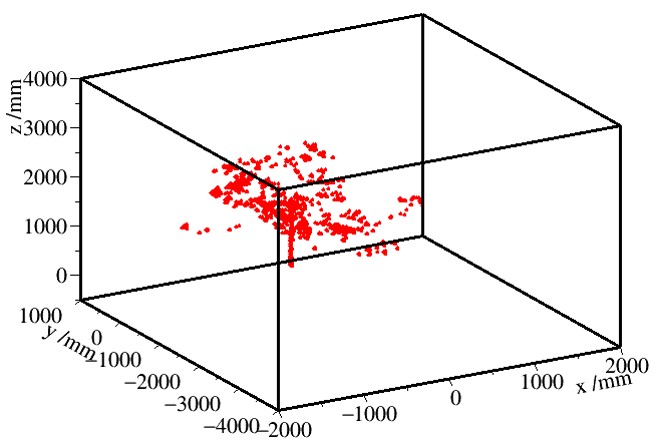}
    \caption{1\,GeV $\pi ^+$}
	\label{fig:sim:particle:muon}       % Give a unique label
	\end{subfigure}	
    \caption{Simulated 1\,GeV particle in liquid scintillator.}
    \label{fig:sim:particle}
\end{figure*}

With further improved sensitivity of the system, it is possible to be used to tag the topology of different particles to do particle identification as simulated in Fig.\,\ref{fig:sim:particle}, where electron, positron, gamma, alpha, proton, pion of 1\,GeV in liquid scintillator can be further identified through their topology for further direction or physics study.

\section{Summary}
\label{sec:1:summary}

With the crystal system viewed by the single photon sensitive camera and PMTs, system calibration was further discussed. The muon track imaging was tested further, and the data were analyzed according to the understanding of the characteristic expectation. Some possible tracks are identified with the averaged signal intensity cuts. But there still are a few critical items that need to be finalized. Some improvements are proposed and suggested. The realization of muon track direct measurement is valuable for future experiments and applications.

\section*{Acknowledgments}
\label{sec:1:acknow}

This work was supported by the National Natural Science Foundation of China (NSFC) No.\,11875282 and 11475205, the State Key Laboratory of Particle Detection and Electronics, SKLPDE-ZZ-202208.

%\appendix
%\section{Some title}
%Please always give a title also for appendices.

%\acknowledgments

%This is the most common positions for acknowledgments. A macro is
%available to maintain the same layout and spelling of the heading.

%\paragraph{Note added.} This is also a good position for notes added
%after the paper has been written.

\bibliographystyle{unsrtnat}
\bibliography{allcites}   % name your BibTeX data base


%\begin{thebibliography}{99}

%\bibitem{a}
%Author, \emph{Title}, \emph{J. Abbrev.} {\bf vol} (year) pg.

%\bibitem{b}
%Author, \emph{Title},
%arxiv:1234.5678.

%\bibitem{c}
%Author, \emph{Title},
%Publisher (year).

% Please avoid comments such as "For a review'', "For some examples",
% "and references therein" or move them in the text. In general,
% please leave only references in the bibliography and move all
% accessory text in footnotes.

% Also, please have only one work for each \bibitem.

%\end{thebibliography}
\end{document}